\documentclass[sigconf]{acmart}
\usepackage{enumitem}

\AtBeginDocument{%
  }

\usepackage{comment}
\usepackage{threeparttable}
\usepackage{todonotes}

\begin{document}

\title{On the Use of Large Language Models for Qualitative Synthesis}

\author{Sebastián Pizard}
\affiliation{%
  \institution{Facultad de Ingeniería, Universidad de la República}
  \city{Montevideo}
  \country{Uruguay}}
\email{spizard@fing.edu.uy}

\author{Ramiro Moreira}
\affiliation{%
  \institution{Facultad de Ingeniería, Universidad de la República}
  \city{Montevideo}
  \country{Uruguay}}
\email{lidnele@hotmail.com}

\author{Federico Galiano}
\affiliation{%
  \institution{Facultad de Ingeniería, Universidad de la República}
  \city{Montevideo}
  \country{Uruguay}}
\email{federicogaliano88@gmail.com}

\author{Ignacio Sastre}
\affiliation{%
  \institution{Facultad de Ingeniería, Universidad de la República}
  \city{Montevideo}
  \country{Uruguay}}
\email{isastre@fing.edu.uy}

\author{Lorena Etcheverry}
\affiliation{%
  \institution{Facultad de Ingeniería, Universidad de la República}
  \city{Montevideo}
  \country{Uruguay}}
\email{lorenae@fing.edu.uy}

\renewcommand{\shortauthors}{Pizard et al.}

\begin{abstract}
Large language models (LLMs) show promise for supporting systematic reviews (SR), even complex tasks such as qualitative synthesis (QS). However, applying them to a stage that is unevenly reported and variably conducted carries important risks: misuse can amplify existing weaknesses and erode confidence in the SR findings. To examine the challenges of using LLMs for QS, we conducted a collaborative autoethnography involving two trials. We evaluated each trial for methodological rigor and practical usefulness, and interpreted the results through a technical lens informed by how LLMs are built and their current limitations. 
\end{abstract}


\maketitle

\section{Introduction}

Research synthesis is the systematic integration and appraisal of knowledge and research findings on a specific question, aiming to improve their generalizability, applicability, and accessibility \cite{Wyborn2018}. By organizing and distilling existing evidence, it helps build cumulative knowledge and facilitates its use \cite{Santos2020}. The term is often used as \textit{"a collective term for a family of methods that are used to summarize, integrate, combine, and compare the findings"} \cite{Cruzes2011b}. Research synthesis may be carried out as the synthesis stage of a systematic review (SR), bringing together results from identified empirical studies, or as a standalone effort whenever a relevant body of empirical work is available \cite{Santos2020}.

Synthesis of empirical research can be quantitative (e.g., meta-analysis to statistically aggregate results) or qualitative (interpreting and integrating non-numerical findings to identify patterns, contrasts, and conceptual insights across studies) \cite{sandelowski2007}. Both are key to organizing and consolidating evidence for use.

Several studies highlight the importance of qualitative synthesis (QS) in software engineering (SE), given the nature of available evidence and the contexts and methods used to generate it \cite{Cruzes2011b, Huang2018, kitchenham2023}. However, qualitative synthesis remains the least well-reported stage of SRs \cite{Cruzes2011b, Huang2018}. For example, Huang et al. were unable to identify the applied synthesis methods in 274 of 328 SRs published between 2005 and 2015 \cite{Huang2018}. More recently, Pizard et al. found that 18 of 23 rapid reviews (i.e., a lightweight SR for resource-constrained settings 
) published before November 2023 did not correctly report their synthesis methods \cite{pizard2025}. As Huang et al. note, this may reflect limited familiarity with QS methods or ad-hoc conduct.

Large language models (LLMs) (i.e., artificial intelligence systems trained on vast text corpora to generate and analyze natural language) offer promising support for the SR process \cite{Thomas2025}. However, applying LLMs to QS, a SR stage that is inconsistently reported and whose conduct is often uncertain, carries significant risks. Without a clear understanding of appropriate uses and associated challenges, LLMs may amplify existing shortcomings and undermine the credibility of SRs findings.

Motivated by these concerns, we conducted this study to better understand the potential of LLMs to support qualitative synthesis. Specifically, we investigate the following research question: \textbf{\emph{RQ: What are the challenges in producing a sound (rigorous and useful) qualitative synthesis using LLMs?}}. To address this question, we conducted a collaborative auto-ethnographic study, involving two trials that utilized LLMs to support different QS methods. We analyzed our trials from two perspectives: methodological rigor and the usefulness of the synthesis. Also, we analyzed them through a technical lens informed by how LLMs are built and recent advances in the technology.

This study makes four contributions: (i) a set of concepts that characterize QS; (ii) identification of LLM properties pertinent to QS support; (iii) criteria for evaluating the soundness of QS stages and their results; and (iv) a report of challenges observed across two trials in which LLMs supported two distinct QS.

The remainder of the paper is organized as follows: Sections \ref{sec_qs} and \ref{sec_llms} provide an overview of QS and LLMs, respectively. Section \ref{sec_methods} describes our research method, and Section \ref{sec_results} presents and discuss our results. Section \ref{sec_trust} analyzes the trustworthiness of our study. Finally, Section \ref{sec_conclu} offers concluding remarks.

\section{Qualitative Synthesis}\label{sec_qs}

Qualitative synthesis, within the context of an SR, is the stage in which findings extracted from individual studies are integrated and interpreted to build a collective understanding, identify common patterns, and generate new interpretations or hypotheses~\cite{Thomas2008}.

QS shares many similarities with qualitative analysis (QA), which involves the systematic interpretation of primary qualitative data (e.g., interview transcripts, observational notes, or documents) to identify patterns, themes, and meanings. They differ in scope and purpose: QA focuses on data from a single study, whereas QS integrates findings across multiple studies to produce broader, more abstract insights. This requires comparing and reconciling results across diverse contexts, research questions, and methods, making synthesis a more integrative and interpretive process. In addition, QS is typically conducted as part of an SR, which places extra demands on transparency, methodological rigor, and reproducibility; standards that generally exceed those for individual~QA.

QS encompasses a broad repertoire of methods, many adapted from QA, including thematic summaries, framework synthesis, thematic synthesis, meta-ethnography, grounded theory, textual narrative synthesis, meta-study, meta-narrative, critical interpretive synthesis, ecological triangulation, content analysis, meta-interpretation, and metasummary (e.g., \cite{Barnett-Page2009, Thomas2012}). Due to the differences in aims and procedures, qualitative research, including QS, is not a singular, unified methodology. This diversity makes it challenging to establish universal quality criteria or provide general recommendations for integrating LLMs into QS.

To navigate this diversity, we outline several aspects that characterize QS\footnote{Based on the framework in~\cite{Barnett-Page2009}, adapted and expanded.} and are relevant when considering the use of LLMs.

\paragraph{Data Transformation: Aggregation vs. Interpretation}

Methodologists often describe synthesis methods along a continuum. At one end are \emph{aggregative} approaches, which summarize and integrate findings across studies with minimal reinterpretation; at the other are \emph{interpretive} (or configurative) approaches, which develop new conceptual understandings by reanalyzing and reinterpreting evidence~\cite{Thomas2012}. Method choice depends on the synthesis question and objectives, the nature of the evidence, and practical constraints such as time and resources~\cite{Snilstveit2012}. Aggregative methods are better suited for hypothesis- or theory-testing questions, while interpretive methods are preferable for exploratory questions and for conceptualizing complex issues. Thin, primarily descriptive data may be addressed via aggregation; richer, more nuanced data typically require interpretive approaches.

\paragraph{Epistemological Position: Idealism vs. Realism}

An important aspect of QS is the researchers' epistemological stance. This can be described as a spectrum from an idealist position, in which reality is construed through multiple, alternative human interpretations, to a realist position, in which reality exists independently of thought and can be directly observed and known \cite{Barnett-Page2009}. In general, idealist stances align with constructivist perspectives, whereas realist stances align with positivist perspectives. Aggregative methods are often, though not always, associated with more realistic positions; interpretive methods are often aligned with more idealist positions. Many methods can be used under different epistemologies. For example, thematic synthesis can be pursued from a realist perspective or a more idealist orientation, akin to thematic analysis~\cite{braun2023b}.

\paragraph{Analytical Orientation}
QS may be conducted inductively or deductively~\cite{Thomas2006b}. Induction emphasizes the discovery of patterns, themes, or concepts from the data without predefined categories; deduction applies existing theories to new data. These orientations are not mutually exclusive; many syntheses blend both, balancing data-driven insights with theory-informed expectations.

\paragraph{Coding}
In various qualitative methods, coding plays a crucial role in both analysis and synthesis. Codes label significant segments of text (such as words, phrases, sentences, or paragraphs) to systematically categorize and interpret information ~\cite{Miles2014}. Coding aids in organizing data and uncovering recurring patterns.

\paragraph{Iteration}
All synthesis approaches involve some iteration, but idealist methods tend to be more iterative in nature. Extensive iteration can conflict with positivist perspectives, which emphasize predefined protocols and minimal midstream changes.

\paragraph{Problematizing the Literature}
Some QS methods involve critically examining the primary studies to uncover underlying assumptions, gaps, or tensions. This approach helps to understand how and in what context knowledge was produced, and can, e.g., offer explanations for the heterogeneity of findings. In contrast, approaches that do not engage in such critique often pose narrower questions and include a more limited range of studies.

\paragraph{Conceptual Innovation}
Synthesis is often described as producing a whole greater than the sum of its parts. Methods differ, however, in the extent to which they transform the data. Some primarily describe and summarize primary studies, translating findings across studies (i.e., identifying concepts in one study and recognizing them in others, even when expressed differently~\cite{Thomas2008}). Others aim to generate new interpretations. Synthesis "involves some degree of conceptual innovation, or employment of concepts not found in the characterization of the parts and a means of creating the whole" \cite{Strike1983}. Not all methods make their transformation strategies explicit, and when primary studies directly address the review question, conceptual innovation may be unnecessary~\cite{Thomas2008}.

\paragraph{Data Heterogeneity}
A central challenge is heterogeneity, particularly in determining what constitutes "findings" \cite{Thomas2008}. Unlike quantitative research, which yields standardized outcomes, qualitative studies report results in varied forms, including raw participant quotes, thematic interpretations, and author conclusions. This variation complicates identification and comparison across studies. True findings must be distinguished from raw data, analytic procedures, and researchers' interpretations, adding complexity to synthesis \cite{sandelowski2007}. Heterogeneity also encompasses methodological diversity~\cite{Barnett-Page2009}. While most methods can accommodate such diversity, they differ in how explicitly they manage comparisons. Approaches with a stronger emphasis on critical appraisal, often those aligned with idealist perspectives, may better address methodological variation and sometimes use it to explain differences in findings. However, their strategies are not always clearly described.

\paragraph{Quality Assessment}
Quality assessment of primary studies in QS methods varies widely, reflecting different epistemological assumptions. More realist approaches often apply structured, predefined criteria that influence inclusion and weighting, treating appraisal as a transparent core of the review. Idealist approaches are more flexible and interpretive, prioritizing relevance or theoretical contribution over standardized checks; judgments may rely on researcher expertise or reflexive assessment rather than formal tools.

\paragraph{Synthetic Product}
Synthesis outputs vary based on their intended use. Methods aligned with positivist traditions often aim for directly applicable findings for policy and practice. Constructivist-oriented approaches tend to produce more conceptual or metaphorical outputs that may require further interpretation before use; their primary contribution is often theoretical development and agenda-setting for future research.

\paragraph{Researcher Subjectivity}
Responsible qualitative work acknowledges researcher subjectivity in ways consistent with the study's epistemological and ontological stance~\cite{Braun2023, 
Tong2007, Tracy2010}. In realist approaches that emphasize coding reliability (e.g., content analysis or certain forms of thematic analysis), subjectivity is treated as a potential source of bias to be measured and controlled. More idealist approaches, such as reflexive thematic analysis, treat subjectivity as inherent and valuable, requiring critical reflection on positionality and influence. These considerations apply equally to QS.

\paragraph{Quality of a QS Stage}
To our knowledge, no work has explicitly defined the characteristics of a sound (well conducted and with relevant results) QS stage. Instead, existing literature primarily focuses on two areas: (i) the validity of qualitative SRs (QSRs) as a whole, for example Sandelowski and Barroso's realist-oriented framework \cite{sandelowski2007}, and (ii) the confidence in individual QSR findings, exemplified by the GRADE-CERQual approach from the Cochrane Qualitative and Implementation Methods Group~\cite{Lewin2018}. Additionally, reviews of QSR assessment checklists are also available \cite{flemming2021}.

\section{Large Language Models}\label{sec_llms}

\paragraph{Language models.}

Language Models (LMs) are machine learning models that predict upcoming words~\cite{jurafsky2025}.
This is done by assigning a probability to each \textit{token} (which denotes a subword, not necessarily a whole word) of a predefined vocabulary.
These probabilities form a probability distribution over the vocabulary.
Tokens with higher probability correspond to more probable continuations for a given input text.
More formally, we note the conditional probability of the next token $w_i$ given the preceding sequence $w_1, w_2, ..., w_{i-1}$, as provided by the language model, as:
$$P_{LM}(w_i|w_1^{i-1})$$

It can be observed that this is equivalent to assigning probabilities to entire text sequences.

\paragraph{Large language models (LLMs).}

We refer to LLMs as deep neural networks with billions of parameters trained for the task of language modeling.
These models typically use variants of the Transformer architecture~\cite{vaswani2017}.
Common variants include the encoder-only Transformer used in models such as BERT~\cite{devlin2019}, which produces contextual word representations, and the decoder-only Transformer used in models such as GPT-4~\cite{openai2024}, which generate text autoregressively, the type used by tools such as ChatGPT\footnote{\url{https://chatgpt.com/}}.

\paragraph{Pretraining, instruction tuning and alignment.}

Modern LLMs are typically trained via a three-stage pipeline:
\textit{(1) pretraining}, in which the model is trained on vast text corpora (on the order of trillions of tokens; e.g., Llama 3 was trained on 15T tokens~\cite{grattafiori2024});
\textit{(2) instruction tuning}, in which the model is fine-tuned on a much smaller corpus of instructions paired with correct responses; \textit{(3) preference alignment}, in which the model is further optimized to align with human preferences, commonly using reinforcement learning methods~\cite{ouyang2022}.

\paragraph{Decoding.}

After completing the training, we have a frozen model optimized for next-token prediction and aligned with human preferences.
The model is then used for \textit{autoregressive} generation: given a \textit{prompt} (the input text), the model generates the next token, appends it to the context, and repeats this process.
An essential step in autoregressive generation is the decoding process,
which specifies how the next token is chosen from the model's probability distribution.
A straightforward option is to select the most probable token (greedy decoding).
This method is deterministic: given a prompt, the resulting generation is always the same.
Popular tools like ChatGPT generally employ sampling-based decoding
, where tokens are sampled according to their likelihood as determined by the LLM's output.
This method yields more diverse outputs but is non-deterministic, which is why these tools can produce different responses given the same prompt.
This behavior can be problematic, as it hinders reproducibility.

\paragraph{Prompting.}

Several prompting methods have been proposed to improve results.
\textit{In-context learning} refers to the emergent ability of these models to learn new tasks from the prompt.
\textit{Zero-shot learning} refers to providing only an instruction for a task to the model, whereas \textit{Few-shot learning} involves adding demonstrations of the solved task to the prompt, which improves model performance~\cite{brown2020}.
\textit{Chain-of-thought}~\cite{wei2022} involves prompting the model to generate a coherent series of intermediate reasoning steps that lead to the final answer, improving results on arithmetic, commonsense, and symbolic reasoning tasks, while improving explainability.

\paragraph{Hallucinations.}

Due to the construction and training of LLMs, these models have inherent limitations that can pose safety and reliability risks.
\textit{Hallucinations} refer to the tendency of models to generate text that is nonsensical or unfaithful to the provided source input~\cite{ji2023}.
Moreover, outputs lack provenance: it is not possible to trace a generated response back to specific training sources.
Methods such as Retrieval-Augmented Generation (RAG) mitigate this issue by retrieving the most relevant documents for the user's question and incorporating them into the prompt~\cite{gao2024}.
These capabilities are already integrated into tools like ChatGPT, with functionalities such as ChatGPT Search\footnote{\url{https://openai.com/index/introducing-chatgpt-search/}} and Deep Research\footnote{\url{https://openai.com/index/introducing-deep-research/}}.
Although such agent-like systems make LLMs more robust, hallucinations remain an inherent risk, and a human-in-the-loop is needed to ensure the correctness of the generated response with respect to the sources.

\paragraph{LLMs as black boxes.}

Tools like ChatGPT are effectively \textit{black boxes}: key details about the underlying models and training data are undisclosed, and deployed systems may change over time. This opacity limits auditability, provenance, and reproducibility, which are essential aspects when using these systems in science~\cite{bender2021, bommasani2022}.

\paragraph{Reasoning}

LLMs trained as described above often perform poorly on tasks that require multi-step reasoning and planning~\cite{valmeekam2022}.
Recently, \textit{reasoning models} have emerged: LLMs post-trained with reinforcement learning to generate extended reasoning traces before producing a final answer.
These models show significant performance improvements on tasks such as mathematics, logical reasoning, and programming.
Still, there is no consensus in the research community on whether these models exhibit true reasoning.
For example, Shojaee et al. examined these models through the lens of problem complexity using controllable puzzle environments~\cite{shojaee2025}, and conclude that they cannot sustain reasoning beyond certain complexity thresholds, sometimes exhibiting less reasoning effort as tasks approach those limits.
Reasoning models are more explainable in the sense that one can inspect the generated traces that led to a response.
However, several works indicate that these traces are not always faithful to the final answer~\cite{chen2025, korbak2025}, and some closed-source systems such as ChatGPT do not expose full reasoning traces.

\section{Methods}\label{sec_methods}

\begin{table*}[htp]
    \centering
    \begin{threeparttable}[b]
    \caption{Characteristics of a Sound Qualitative Synthesis\tnote{a}, distilled from~\cite{sandelowski2007, Lewin2018, krefting1991, flemming2021}}
    \label{tab_char}
    \footnotesize
\begin{tabular}{p{5.2cm}p{12.2cm}} \toprule
Characteristic                                                                                                                                                                                                     &  Description    \\ \midrule
C1. The findings are grounded in data from the primary studies and account for the observed patterns~\cite{sandelowski2007, Lewin2018, krefting1991, flemming2021}. They provide a full and fair representation of the perspectives and understandings of the primary study authors~\cite{sandelowski2007, krefting1991}. & Findings are grounded in primary studies, with clear attribution of which studies contribute to each finding. Each finding should provide a convincing explanation of the observed patterns. The findings seek to accurately reflect the perspectives and interpretations of the primary study authors. Reviewers should also report and attempt to explain contradictory evidence from primary studies that may complicate or challenge their main conclusions. \\ \hline

C2. The findings have practical relevance to the research question, and information is available to support their transferability to other contexts~\cite{sandelowski2007, Lewin2018, krefting1991}. & The findings are relevant and timely for the research questions and for the intended use of the QS outputs, bringing forward distinct themes and concepts that deepen understanding. They also provide insights into how these findings can be translated into practical applications and applied across various contexts. For instance, by incorporating details about the settings of the primary studies in which the findings were identified.  \\ \hline

C3. The findings can be appropriately weighted and considered for use, which may require identifying and reporting the methodological limitations of the primary studies that underpin them~\cite{Lewin2018, flemming2021}. & When appropriate, depending on the type of secondary study in which the synthesis is embedded, and to adequately report confidence in the findings, the methodological limitations of the primary studies are assessed, and potential biases (e.g., dissemination bias) are discussed. Generally, a separate assessment of the methodological strengths and weaknesses of the primary studies may be required. \\ \hline

C4. The choice of synthesis method is appropriate and well justified by the reviewers~\cite{sandelowski2007, krefting1991, flemming2021}.    & The authors select a synthesis method that is consistent with the research question, the chosen epistemological approach, and the available primary study data. They also provide a clear and well-founded justification for this choice.   \\ \hline

C5. The conduct of the method is adequate, its reporting is transparent, and sufficient information is provided to enable auditing~\cite{sandelowski2007, krefting1991, flemming2021}.  & The authors provide information on the conduct of the synthesis method that demonstrates it was applied appropriately. Furthermore, the reported details make it possible to audit the process. For example, by including intermediate outputs.   \\ \hline

C6. The authors address their subjectivity appropriately and in alignment with the epistemological approach they have chosen~\cite {krefting1991}.       & In realist approaches, addressing subjectivity typically involves independent coding by multiple researchers, followed by the calculation of agreement metrics. In contrast, more idealist (nonpositivist) approaches require authors to demonstrate reflexivity about the interpretive process. This includes critically examining potential influences, such as their background, perspectives, positions, and interests (especially regarding the research topic), as well as biases introduced by the software tools used \cite{Weitzman2003}.    \\ \bottomrule
\end{tabular}
\begin{tablenotes}
\item [a] These characteristics should be met whether or not LLMs or other technologies are used to support the QS stage.
\end{tablenotes}
\end{threeparttable}
\end{table*}

To answer our research question, we conducted \textbf{two exploratory trials} that followed a similar approach (hereafter, 1st and 2nd Trial). In both cases, Moreira and Galiano worked independently under Pizard's supervision, meeting weekly to monitor progress, review results, and discuss alternatives. Each trial comprised three stages: first, a piloting stage in which they experimented with LLMs on a QS task comparable to the target QS and produced a plan that included one or more prompting strategies; second, application of this plan to synthesize the final set of primary studies; and third, evaluation of the resulting synthesis against predefined criteria. In particular, and because we found no prior work clarifying this aspect, we distilled a set of characteristics of a sound QS stage (Table~\ref{tab_char}), informed by prior work on QSR quality (Section~\ref{sec_qs}) and the trustworthiness framework for qualitative research \cite{krefting1991}. To ensure a realistic training environment, Moreira and Galiano had the freedom to utilize LLMs as needed during their QS. They later validated their strategies with Pizard. In both cases, they chose not to perform a prior data-extraction stage and instead provided the full articles to the LLMs. Both trials were based on an unpublished manual synthesis to prevent potential data leakage, i.e., the risk that LLMs might have previously encountered the manual synthesis and, as a result, could overfit their responses. 

Our \textbf{research method} is based on retrospective trials, with collaborative autoethnography to evaluate and synthesize the problems and risks observed during the trials. Autoethnography is a qualitative approach in which researchers systematically analyze their own experiences, interactions, and reflections as data \cite{Cunningham2005, Sharp2016}. It has been used to study the use of LLMs for QA \cite{Ferguson2025, AlFattal2025} and is well suited to complex inquiries like ours. It enabled us to treat our research practice as data. We assembled multiple data sources, including video recordings of all meetings, meeting notes, full logs of our interactions with LLMs and other GenAI tools, and a personal diary kept by the first author that documented key decisions. 

For \textbf{data analysis}, we transcribed the recordings, consolidated the notes, and engaged in iterative reflection and theme development, following a reflexive thematic analysis approach \cite{braun2021}. To detect the challenges we faced, we applied a mixed approach. First, we conducted an inductive analysis, complemented by a deductive analysis based on three sources: (1) challenges in using LLMs for QA, derived initially from \cite{Barros2025, Leca2025} and updated with insights from the 12 studies about using LLMs for QA analyzed in the 2nd Trial; (2) challenges of using LLMs to conduct SRs identified by \cite{Felizardo2025} applicable to QS; and (3) challenges related to the use of LLMs in science identified by \cite{Binz2025}. Finally, we discussed our results and reflections with Sastre (specialized in GenAI and LLMs) and Etcheverry (specialized in data science and data bias), who were not involved in the trials. Their feedback helped to place our findings within the broader literature and provided an external, reflective perspective.

About our \textbf{background and positionality}, which should be identified and reported in qualitative research \cite{Berger2015}, Pizard holds a PhD in Computer Science focused on the adoption of Evidence-Based Software Engineering (EBSE)\footnote{This is the application of evidence-based practice within software engineering.} and has taught an EBSE and SRs course since 2017 \cite{pizard2021}. He also led several QS involving the specific methods applied in these trials \cite{pizard2021, pizard2023, pizard2025}. Moreira and Galiano, students in the 2024 EBSE and SRs course, each completed at least one QS as part of their coursework. Because they are nearing completion of a five-year undergraduate program, we consider their background comparable to those of postgraduate students working under supervision. All researchers, especially the trialists, initially held a neutral stance on using LLMs for QS.

\begin{table}[htp]
    \centering
    \begin{threeparttable}[b]
    \caption{Trials Info and QS Characteristics Evaluation\tnote{a}}
    \label{tab_trials}
    \footnotesize
\begin{tabular}{p{\columnwidth}} \toprule
\multicolumn{1}{c}{1st Trial} \\ \midrule
\textbf{Description.} \textit{RQ:} What main issues, challenges, and facilitators do proponents of EBSE report in two foundational position papers? \textit{Method:} Thematic Synthesis (TS). \textit{Manual analysis:} Previously performed by Pizard in collaboration with Barbara Kitchenham. \textit{Process:} Galiano used ChatGPT-4o (Plus suscription) with two prompting strategies (zero-shot and expert persona) and prompt chaining, requesting each analysis step separately. Piloting stage used data and results from an SR on EBSE training \cite{pizard2021}. He also tried a few-shot approach unsuccessfully, as ChatGPT failed to apply TS properly, and he tested Elicit\tnote{b}, which did not support running a QS on a specific set of papers. No prior data extraction was performed. \textit{QS key aspects:} Intermediate between aggregative and interpretive, realist and inductive, and grounded in coding.\\                                                                                          

\begin{tabular}[t]{@{}p{\columnwidth}@{}}\textbf{Results of Evaluation.} In this case, the evaluation of characteristics involved assessing C1, C2, C5, and C6. We present the QS evaluation for each prompting strategy.

\begin{itemize}[leftmargin=0.3cm]
    \item[-] \textit{Zero-shot} (C1) Final outputs were traceable to data, but intermediate steps weren’t always grounded. (C2) Themes were relevant but overly superficial; researcher felt more manual refinement was needed. (C5) Step-by-step made the procedure clear, yet there were doubts about correct method application. (C6) LLM's stated "position" was neutral/superficial; limitations were more detailed but implicitly positivist and focused only in dataset. 
    \item[-] \textit{Expert persona.} (C1) Many themes grounded, but several findings weakly supported or speculative. (C2) Themes felt less relevant/coherent; lacking expected coherence. (C5) Same concerns as above; additionally, the LLM sometimes took undisclosed actions not queried. (C6) Expert "positions" read as plausible clichés consistent with roles; limitations remained high-level. 
\end{itemize} \end{tabular}  \\  \midrule

\multicolumn{1}{c}{2nd Trial} \\ \midrule

\textbf{Description.} \textit{Aim:} Synthesize evaluations of LLMs used to support QA, extending \cite{Barros2025}. The synthesis included summaries as a quality control check.  \textit{RQs:} RQa: How was the effectiveness of the LLM evaluated, and what were the main outcomes? RQb: What limitations were reported? \textit{Dataset:} Pizard retrieved 12 additional studies using SCOPUS. \textit{Method:} RQa was addressed through categorizations, and RQb through deductive content analysis (using the topics identified in the original MS), complemented with inductive analysis. \textit{Manual analysis:} Conducted by Pizard. \textit{Process}: Galiano used ChatGPT 5 Thinking (Plus suscription) and Gemini 2.5 Pro with zero-shot prompting. Piloting stage used the original MS dataset and results \cite{Barros2025}. No prior data extraction was performed. \textit{QS key aspects:} Aggregative (RQa) and interpretive (RQb), realist, and primarily deductive. \\   

\begin{tabular}[t]{@{}p{\columnwidth}@{}}
\textbf{Results of Evaluation. } We assessed C1, C2, and C5, excluding C6 owing to the predominantly descriptive, realist orientation of the QS methods. The QS evaluation is presented by research question, each employing distinct QS methods. Results are not disaggregated by LLM, as both models exhibited comparable performance.
\begin{itemize}[leftmargin=0.3cm]
    \item[-] Paper \textit{summaries} were very good. 
    \item[-] \textit{RQa (Aggregative/Categorizations):} Outputs were (C1) relevant and (C2) data-grounded; one major discrepancy stemmed from a manually added category value by Pizard, which the LLMs failed to identify. (C5) Asking for citations/justifications improved traceability, though these were sometimes missing initially. 
    \item[-] \textit{RQb (Interpretative/Content Analysis):} This RQ required identifying findings reported as research results in each paper. (C1 \& C2) The initial outputs listed points merely mentioned, not actual results; after restricting the request to results only, responses became less speculative. Both models showed high precision but weak MCC\tnote{c} (just above chance) because they often failed to assign predefined categories. (C5) The “results-only” instruction was not always followed, and without prompt chaining it is unclear whether content analysis was consistently applied.
\end{itemize}
\end{tabular}
\\ \bottomrule
\end{tabular}
\begin{tablenotes}
\item [a] C3 was out of scope (with a dedicated research line, e.g., \cite{Thelwall2025, Honghao2024}), and C4 was not assessed because the QS methods were predetermined by the reference QSs.
\item [b] Elicit is an AI research assistant that searches, summarizes, and extracts data from academic papers to automate parts of literature reviews and SRs. https://elicit.com/
\item [c] Matthews correlation coefficient (MCC) quantifies how well predictions agree with the ground truth, using true positives, true negatives, false positives, and false negatives. It is preferred in this setting because it remains robust under class imbalance.
\end{tablenotes}
\end{threeparttable}
\end{table}

\section{Findings \& Discussion}\label{sec_results}

To contextualize our findings, Table \ref{tab_trials} presents each trial and the main results from evaluating the QSs against the characteristics of a sound QS (refer to Table \ref{tab_char})\footnote{We used the manual syntheses as the reference baseline. For relevance (C2), we also considered the intended use of the outputs. We treated the LLM-assisted synthesis as a stand-alone alternative to the manual synthesis, not a complement.}.

The main challenges we faced in using LLMs to support QS are\footnote{Due to space limitations, traceability information to the underlying data is reported in the supplementary material (\href{https://doi.org/10.5281/zenodo.17385700}{https://doi.org/10.5281/zenodo.17385700}), along with complete traces of the LLM interactions during the trials.}:

\textbf{1. Accounting for the Nuances of Qualitative Synthesis.} Our first challenge was to properly identify the nuances of QS so that tools like LLMs could be used appropriately within its process. Treating qualitative methods only at a superficial conceptual level (a stance we observed in several studies on using LLMs for QA and also confronted at the start of our own trials) seems not only naïve but risky. To address this, we elaborated a set of QS relevant aspects (refer to Section 2) that we used to design and evaluate our trials. 

\textbf{2. Understanding LLM Capabilities and Limits.} While conducting the 1st Trial, we confronted a second challenge: understanding the complexity of LLMs (their architecture, training regime, and inference behavior) and, consequently, what they can and cannot be used for. As with QS, we observed that many studies treat this technology superficially; e.g., several empirical papers evaluating LLMs for QA do not even mention hallucinations, a key limitation of LLM-based approaches. To address this challenge, we invited two specialists to the team: Sastre (LLMs and generative AI) and Etcheverry (data science and data bias). Sastre helped us articulate key LLM concepts and clarify their capabilities and limitations (see Section 3), while Etcheverry focused on data considerations and bias. Their participation enabled a more critical interpretation of the process and results of our trials.

\textbf{3. Difficulty in Evaluating Qualitative Syntheses.} In evaluating the 1st Trial, we found that, unlike other stages of an SR, direct comparisons between a manual synthesis and a LLM-assisted one are not always straightforward or especially informative. Assessing qualitative syntheses (both their processes and their products) seems to be inherently challenging, particularly given the wide variety of available methods and the multiple aspects involved in this activity (see Section 2). As we mentioned earlier, to address this challenge we distilled a set of characteristics for a sound QS (see Table~\ref{tab_char}). We then used this list to assess both trials.

\textbf{4. Sensitivity to Prompts.} LLMs can produce markedly different outputs from small wording changes, or even using the same prompt, which can introduce bias, undermine traceability, and hinder replication. To mitigate this sensitivity, that we early observed in the 1st Trial, we tested and refined prompts on a separate problem and dataset before conducting the QS, and we tried to use explicit, standardized instructions.

\textbf{5. Possibility of Incorrect Responses.} LLMs are not designed to strictly follow user instructions and are prone to hallucinations. Although repeated prompt refinement allowed us to obtain responses grounded in the data, this sometimes reduced relevance, as meaningful findings often require interpretation beyond simple description. For example, in one test of the 1st Trial, instructing the model to emulate a consensus among multiple experts led it to misinterpret several arguments from the primary studies, substantially distorting their meaning.

\textbf{6. Lack of Transparency.} Because their architecture is optimized to produce the most probable continuation rather than to follow instructions exactly, LLM behavior is not guaranteed to align with the method they are instructed to follow. Throughout our trials, we often doubted that the models were applying the specified synthesis method. In particular, when our instructions were not sufficiently complete and precise, their outputs did not appear to follow the method’s required steps. Providing rich, precise context is therefore essential, especially for less-used methods that are likely underrepresented in LLM training data. We also found that prompt chaining helped validate intermediate results and yielded more consistent outcomes. Finally, for more iterative QSs, additional effort should be made to preserve an auditable trace (e.g., storing all prompts and intermediate/final responses).

\textbf{7. Lack of Interpretation.} LLMs do not truly interpret information; they generate the most probable responses. LLMs may be more appropriate for purely descriptive syntheses. When given less realist assignments (see challenge 6), we observed superficiality and misinterpretations. They also failed to mirror reflective decisions present in the manual synthesis (e.g., adding a new category after early classifications). Hence, LLMs seem poor fit for tasks requiring conceptual innovation. They may be ill-suited to problematizing the literature or evaluating the quality of primary studies. Finally, data heterogeneity proved to be a major challenge in the 2nd Trial. The models' syntheses often included findings from related work rather than focusing solely on the results of the primary studies. When we explicitly asked them to report only primary studies results, their responses became much shorter and were not always based on primary studies results. This issue likely stemmed both from the models' limited interpretive capacity and from the fact that, in the papers, results were reported in specific sections that the models did not consistently recognize.

\textbf{8. Presence of Inherent Biases}
Because of how LLMs are built, they carry biases introduced during training and fine-tuning. In particular, current LLMs primarily reflect WEIRD populations (Western, Educated, Industrialized, Rich, and Democratic) and cannot easily be prompted to represent non-WEIRD populations \cite{
Binz2025}.

In the 1st Trial, we requested explanation of the LLM's position and the limitations of its QS, the responses were often superficial or cliché, though they did reveal some relevant biases. For instance, when asked to produce realistic syntheses, LLMs appeared to default to a positivist stance. As LLMs generate probabilistic outputs rather than reflecting on their process, they tend to provide generic descriptions rather than accurate accounts of their own actions. A better strategy could be to request control information in real time at each analysis step, rather than retrospectively.

Inherent biases are likely to have a stronger effect on more idealistic syntheses. But researchers should always validate the interpretations produced by LLMs; not only because they remain ultimately accountable for the research results but also for humanistic and philosophical reasons. In this regard, we agree with Binz et al. \cite{Binz2025}, who ask, "Whose understanding matters?" and argue that human understanding should remain a core goal of science.

\textbf{9. Reproducibility Issues.} Ensuring reproducibility in science is critical, particularly when conducting SRs. Commercial LLMs are black boxes, and providers may modify their models (sometimes without notifying users) \cite{Binz2025}. In our trials, we did not explicitly attempt to reproduce identical prompts, but we observed that responses were not always consistent across runs. As \cite{Binz2025} notes, one way to address reproducibility concerns is to use open-source models. However, these currently require substantial learning, installation effort, and technical expertise, which makes them less accessible and limits their widespread use.

\textbf{10. LLMs Are Not Purpose-Built for Tasks Like Qualitative Synthesis.} Based on challenges 4 through 9 and Sections 2 \& 3, we identified another challenge related to the nature of LLMs and the characteristics of QS: LLMs were not originally designed to address the types of problems that arise in QS. Despite this, they are highly flexible, and continual improvements and extensions to their base architectures increasingly enable support for such activities. Risk-management processes and a human-in-the-loop approach are essential for integrating these tools into QS. The former ensures explicit identification, assessment, and mitigation of risks in how the technology is used; for example, Google reports applying a risk-management approach to LLM-based text summarization in healthcare \cite{Obika2024}. The latter ensures that the researcher leads the synthesis, validates model outputs, and remains accountable for decisions throughout the process \cite{Shah2025}.

\textbf{11. Evaluating QS Quality Requires Significant Effort.} Assessing whether a QS result is sound (i.e., meets the characteristics in Table \ref{tab_char}) is labor-intensive for both manual and LLM-assisted syntheses. In the latter case, the effort is even greater: LLMs tend to be more verbose than human researchers, and we frequently had to request corrections or clarifications, which meant reviewing a larger volume of material.

\textbf{12. Vulnerability to Confusing Form with Content.} In our trials, we often found ourselves perceiving the results as useful at first glance, only to discover upon closer examination that they were not. As \cite{Binz2025} notes, we are prone to mistaking LLM output for meaningful or reliable information because our linguistic processing is instinctive and reflexive. In other words, we are poorly equipped to critically evaluate LLM output because we naturally try to make sense of it. This poses a significant challenge when there are no appropriate tools for assessing the technology's use. A good practice could be to establish clear quality criteria and verification methods before using LLMs \cite{Binz2025}, our list of characteristics for a sound QS can serve this purpose.

\textbf{13. LLMs as a Compelling Innovation.} LLMs are an appealing innovation to adopt, which can make researchers more inclined to use them, a tendency we also experimented in our trials. According to Rogers’ diffusion of innovations theory \cite{rogers2003}, potential adopters assess innovations by compatibility, trialability, relative advantage, observability, and simplicity. In our context, compatibility appears high because the values and biases embedded in LLMs are largely invisible, so researchers may view them as harmless tools aligned with the scientific method even when that is not the case. Trialability is high because LLMs are easy to try and offer capable free versions. Relative advantage seems clear because LLMs can produce useful results much faster than manual synthesis. Observability is strong because their outputs often appear useful and trustworthy at first glance. Simplicity is salient because, as noted by \cite{Binz2025}, LLMs operate as general-purpose tools that reduce the need for specialized solutions and shorten the learning curve. Taken together, these factors make LLMs particularly compelling and can encourage their use even when limitations are not fully considered.

\section{Trustworthiness of this study}\label{sec_trust}

We reviewed the weaknesses of our work using Krefting's criteria \cite{krefting1991} from Guba's trustworthiness framework \cite{guba1981}. 

\textit{Credibility:} the accuracy and plausibility of findings, supported by sound methods and reflexive practice. Our trials were conducted by two junior researchers; to mitigate potential deviations, Pizard led the both trials and performed the initial evaluations of the outcomes. We also worked with specialists in LLMs and data science to deepen our understanding of these topics and to provide an independent appraisal of our procedures and results.
\textit{Transferability:} the extent to which results may apply in other contexts, given rich description of setting and atypical factors. We report only challenges we faced 
; some likely issues remained unconfirmed, e.g., LLM usage costs, token-window limits, and copyright constraints when handling papers (we assumed opting out of model training on our data was sufficient). Also, we did not perform a distinct data-extraction phase, nor did we include two characteristics of QS in our study: (C3) weighting results according to the quality of the primary studies and (C4) the choice and justification of the synthesis method.
\textit{Confirmability:} the traceability of conclusions to the data through a documented, auditable analysis. To maximize transparency and auditability we have produced detailed records of all our activities stored in our supplementary materials.
\textit{Dependability:} the findings are consistent with the raw data and auditable. Sastre and Etcheverry provided expertise to assess the validity of the trials. Also, the links between data and our findings are documented in the supplementary materials.

\section{Concluding remarks}\label{sec_conclu}

The {Responsible AI in Evidence SynthEsis (RAISE) guidelines~\cite{Thomas2025}} require that AI use in the SR process be justified, methodologically sound, context-appropriate, and not compromise trustworthiness. Our results show serious challenges to using LLMs for QS while meeting these conditions: they can produce incorrect outputs, lack transparency about processes and biases, deliver plausibility not interpretation, exhibit inherent biases, and hinder reproducibility. Their elaborate, plausible answers and their appeal as an attractive innovation do not help, since we are prone to consider them as useful even when they may not be. LLMs may help with narrow tasks such as paper summaries or descriptive categorizations, but verifying that an LLM-supported QS is sound may take as much or more effort as doing it manually.

\begin{table}[htp]
    \centering
    \caption{Recommendations for using LLMs for QS}
    \label{tab_recom}
    \footnotesize
\begin{tabular}{p{\columnwidth}} \toprule

\begin{enumerate}[leftmargin=0.3cm]
    \item Identify the characteristics of the QS to be conducted and plan an appropriate use of LLMs. For example, for an interpretive QS, verify that any interpretations produced by LLMs align with your positions.
    \item Study the QS methods to be used thoroughly so you can guide and validate what the LLMs produce.
    \item Implement risk management processes with a human-in-the-loop approach. State the QS purpose and potential LLM harms explicitly so you can mitigate them, and ensure researchers lead the QS and validate all intermediate and final outputs.
    \item Plan and take specific actions to address LLM-specific issues for these kind of tasks (e.g., incorrect answers, inherent biases, lack of transparency).
    \item Apply predefined criteria to evaluate the QS and its results. The characteristics of a sound QS proposed in this work may be useful here. Note that validation may require substantial effort.
    \item Finally, as in all qualitative research \cite{Lenberg2024}, emphasize reflexivity and provide an auditable report of the process and its results.
\end{enumerate} \\
\bottomrule
\end{tabular}
\end{table}

However, LLMs will likely be used to support QS; in these cases, proceed with great caution. Table \ref{tab_recom} outlines practical recommendations. As the technology continues to improve, further research is needed that covers QS with diverse characteristics and intended uses, and that evaluates open-source alternatives or LLM-based tools specifically designed to support QS. \newline
\newline \textbf{Acknowledgment}
\newline  We thank Prof. Barbara Kitchenham for her comments on the initial plan for this research and for reviewing drafts of this paper.

\bibliographystyle{ACM-Reference-Format}
\bibliography{main}

\end{document}